\documentclass[conference]{IEEEtran}
\IEEEoverridecommandlockouts
\usepackage{cite}
\usepackage[top=0.7in, bottom=1in, left=1in, right=1in]{geometry}

\usepackage{amsmath,amssymb,amsfonts}
\usepackage{algorithm}
\usepackage{algorithmicx}
\usepackage{algpseudocode}
\usepackage{graphicx,caption,subcaption}
\usepackage{textcomp}
\usepackage{xcolor}
\def\BibTeX{{\rm B\kern-.05em{\sc i\kern-.025em b}\kern-.08em
    T\kern-.1667em\lower.7ex\hbox{E}\kern-.125emX}}
\begin{document}

\title{Power Allocation and RIS Elements Optimisation for Reconfigurable Intelligent Surfaces assisted RSMA

}
\author{Abdullah Qayyum,~and Maziar Nekovee* \\
\textit{6G Lab, School of Engineering and Informatics}\\
\textit{University of Sussex,}, United Kingdom \\
Email:\{A.qayyum,~M.nekovee\}@sussex.ac.uk
}

\maketitle

\begin{abstract}
This paper proposes power allocation and the number of reconfigurable intelligent surfaces (RIS) elements optimisation in a RIS-assisted rate splitting multiple access (RSMA) system. The optimised RIS-RSMA (ORIS-RSMA) method determines the optimal number of RIS elements and the power allocation factors for both common and private parts of a message. Additionally, it maximises the sum rate while ensuring that a target common rate is satisfied. The performance of the proposed ORIS-RSMA is compared to that of the conventional RIS-RSMA and RSMA. Simulation results show that ORIS-RSMA achieves a higher sum rate.
\end{abstract}


\section{Introduction}
With 5G already in commercial use, there is growing demand for 6G networks to support emerging technologies such as autonomous cars, holographic communication, augmented reality and virtual reality. These emerging technologies require high data rates and reliability. As a result, researchers are working towards the next generation of wireless communication, which is 6G. In 6G, the focus is towards the fully connected world where everything from machines to humans will be connected to the internet with a peak data rate of up to 1 Tbps \cite{gui20206g}. New multiple access (MA) techniques are explored to achieve this high data rate. 

Rate splitting multiple access (RSMA) is a non-orthogonal transmission scheme considered for the next generation of wireless communication \cite{mao2022rate}. RSMA refers to the class of multiple access schemes based on the rate splitting (RS) principle, which dates back to the early era of the internet \cite{han1981new, 1055812}. The rate splitting concept (RS) divides a user message into two or multiple parts so that single or multiple receivers can decode the message. This approach allows for more flexible interference management. In space division multiple access (SDMA), interference is treated as noise; in non-orthogonal multiple access (NOMA), interference is decoded completely. In contrast, RS treats some interference as noise while decoding the remaining part, thereby unifying the two extremes. The interest in the RS increased in the past decade as the research for a multiple access (MA) scheme capable of managing interference effectively in the presence of imperfect channel state information (CSI) boomed \cite{davoodi2016aligned}. Other advantages of RSMA include enhanced spectral efficiency, low latency compared to other multiple access schemes and robustness to mobility. One of the key challenges in RSMA is the common rate constraint. This is the minimum rate at which all users must be able to decode the common message. The user with the weakest channel typically determines this rate, which limits the overall system performance..

Reconfigurable intelligent surfaces (RIS), also known as intelligent reflecting surfaces (IRS), are arrays of metasurfaces. These surfaces can modify the wireless channel by modifying the phase of the incident signal. As a result, RIS has become a significant area of focus in 6G \cite{liu2021reconfigurable}. Being a new topic in wireless communication, researchers are working to explore the full potential of RIS as it can mitigate various challenges in wireless communication systems \cite{dai2021wireless, di2020reconfigurable}. The benefits of RIS include its easy deployment, cost-effectiveness, increased spectral efficiency by mitigating interference and energy efficiency. These advantages motivate researchers to combine RIS with other technologies, such as rate splitting multiple access (RSMA).

The combination of RIS and RSMA can result in several mutual gains such as less complexity, enhanced energy \cite{katwe2022rate} and spectral efficiency \cite{katwe2023improved}. These benefits are discussed in detail in \cite{li2022rate}. Additionally, the recent literature focuses on the maximisation of weighted sum rate and minimum rate by optimising RIS phase shift matrices, power allocation and beamforming vectors \cite{singh2023rsma, li2023robust, ge2023rate, huang2023rate}. However, the effect of the optimal number of RIS elements on the performance of RIS-RSMA needs to be investigated, especially when RIS has a large number of reflecting elements \cite{kammoun2020asymptotic}. 

Motivated by this, this paper proposes jointly optimising power allocation and the RIS elements for a downlink RIS-assisted RSMA system. Unlike existing literature, the proposed method optimises the power allocation in RSMA and the allocation of RIS elements for phase shifting to each user. By ensuring Quality of Service (QoS) for the common rate, this optimisation mitigates the impact of the weakest user on the common rate constraint, leading to improved overall sum rate performance. To solve this optimisation problem, an algorithm, Optimised RIS-RSMA (ORIS-RSMA), is formulated, and the algorithm ORIS-RSMA is proposed to solve the optimisation problem. Additionally, the results of the proposed ORIS-RSMA are compared with conventional RSMA and RIS-RSMA to show the performance gains in terms of the sum rate.  

\par
\par
\par
The rest of the paper is organized as follows. Section II describes the system model of RIS-RSMA and the problem formulation. In section III, the proposed optimisation solution is discussed. Section IV discusses the sum rate of the proposed scheme, which is analysed and compared to the conventional schemes. Finally, the paper is concluded in section V.  

\section{System Model and Problem Formulation}
\begin{figure}[t!]
	\centerline{\includegraphics[scale=0.4, trim={1.8cm 0.2cm 0.2cm 0.2cm},clip]{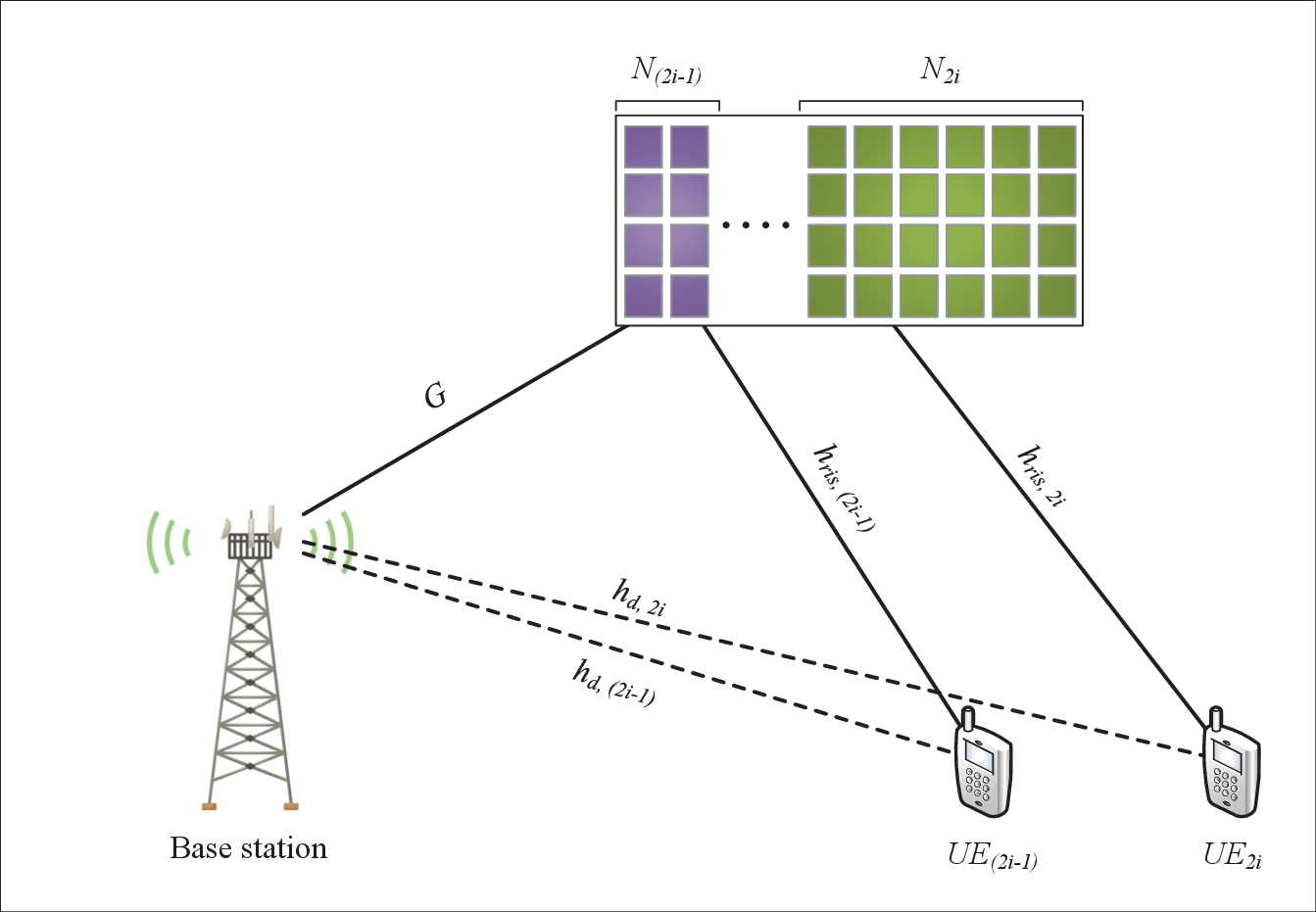}}
	\caption{System model of the proposed RIS-RSMA system.}
	\label{fig1}
\end{figure}
\subsection{RIS-RSMA System}
As shown in Fig. 1, a downlink SISO-1L-RSMA system having a single antenna base station (BS) serving $K$ single-antenna user equipment (UEs) is considered. UEs are represented as $UE_k$ where $k \in \{1, 2, \dots, 2i-1, 2i, \dots, 2I-1, 2I\}$. For ease of exposition, 2 UEs are considered in a group. The number of UEs in a group can be varied. These UEs are served via the $N$ number of a RIS array. Additionally, the total reflecting elements $N$ are partitioned into different segments or sub-surfaces, with each segment containing $N_k$ $\subset$ $N$ elements \cite{zheng2020intelligent} as depicted in Fig.~\ref{fig1}. The RIS is placed to ensure a direct line-of-sight (LOS) with the BS to facilitate signal transmission from the BS to
$K$ UEs. The transmission channels between BS and RIS, BS and UEs, and RIS and UEs are $\textbf{G}$, $\textbf{h}_{d, k}$, and $\textbf{h}_{ris, k}$, respectively. According to the RSMA principle, the $i$-th user group $UE_{2i-1}$ and $UE_{2i}$ data is divided into private and common parts as $\left\{W_{2 i-1}^{\mathrm{c}}, W_{2 i-1}^{\mathrm{p}}\right\}$ and $\left\{W_{2 i}^{\mathrm{c}}, W_{2 i}^{\mathrm{p}}\right\}$, respectively. The common parts of both UEs data $W_{2 i-1}^{\mathrm{c}}$ and $W_{2 i}^{\mathrm{c}}$ are encoded into the transmit signal $s_{2i, 2i-1}^{\mathrm{c}}$ and will be decoded by all UEs. While the private parts are encoded separately into $s_{2i-1}^{\mathrm{p}}$ and $s_{2i}^{\mathrm{p}}$ which will be decoded by their respective UEs. Hence, the transmit RSMA signal to the $i$-th user group $UE_{2i-1}$ and $UE_{2i}$ is $x_k=\alpha_c P_T s_{2 i-1,2 i}^{\mathrm{c}}+\alpha_{p, 2i-1} P_T {s}_{2i-1}^{\mathrm{p}}$ $+$  $\alpha_{p, 2i} P_T {s}_{2 i}^{\mathrm{p}}$  where $\alpha_c$, $\alpha_{p, 2 i-1}$ and $\alpha_{p, 2 i}$ are the power allocation factors for common part, private part of $UE_{2i-1}$ and $UE_{2i}$, respectively \cite{chen2022performance, aswini2024performance}. Also, $\alpha_{p, 2 i-1}$= $\alpha_{p, 2 i}$ = $(P_T- \alpha_c P_T)/K$ where $P_T$ is the total transmit power and $K$ is the total number of user. The signal received at $UE_{k}$can be expressed as:

\begin{equation}
y_{k}=\left(\mathbf{G} \mathbf{\Theta}_{k} \mathbf{h}_{ris, k)}+\mathbf{h}_{d, k}\right) x_k+z_{k},
\end{equation}

where $\mathbf{\Theta}_k=\operatorname{diag}\left(e^{j \theta_1}, \ldots, e^{j \theta_{N_k}}\right) \in \mathbb{C}^{N_k \times N_k}$ is the phases shift matrix of RIS with $\theta_j \in[0,2 \pi], j \in$ $\left\{1, \ldots, N_k\right\}$, and $z_k \sim \mathcal{C N}\left(0, \sigma_w^2\right)$ is the additive white Gaussian (AWGN) noise.

\subsection{RIS-RSMA rates}
According to the RSMA principle, users first decode common parts while considering private parts as noise. Hence, the rate for the common part at $UE_k$ can be represented as: 

\begin{equation}
R_{c,k} = \log_{2} \left (1+ \frac{\alpha_c P_T \mathbf{h_k}}{\alpha_{p, 2i} P_T \mathbf{h_k} + \alpha_{p, 2i-1} P_T \mathbf {h_k} + N_0}\right),
\end{equation}

where $\mathbf{h_k}= \left(\mathbf{G} \Theta_{k} \mathbf{h}_{ris, k)}+\mathbf{h}_{d, k}\right)$ and $N_0$ is the variance of the additive white Gaussian noise (AWGN). To guarantee that each user successfully decodes the common part, the common rate is constrained by $R_{\mathrm{c}}=\min (R_{\mathrm{c}, k})$. For the decoding of the private part, each user performs successive interference cancellation (SIC)\footnote{As 1L-RSMA is considered each user performs SIC only once} to remove the common part from the overall message and decode its private part. Hence, the rate for decoding private parts of $UE_{2_{i}-1}$ and $UE_{2i}$ can be calculated as:

\begin{equation}
R_{p ,2i-1} =\log_{2} \left (1+  \frac{\alpha_{p, 2i-1} P_T \mathbf{h}_{2i-1}}{\alpha_{p, 2i} P_T \mathbf{h}_{2i-1} + N_0}\right),
\end{equation}

\begin{equation}
R_{p ,2i} = \log_{2} \left (1+   \frac{\alpha_{p, 2i}  \mathbf{h}_{2i}}{\alpha_{p, 2i-1}  \mathbf{h}_{2i} + N_0} \right),
\end{equation}

where $\mathbf{h}_{2i-1}= \left(\mathbf{G} \Theta_{2i-1} \mathbf{h}_{ris, 2i-1)}+\mathbf{h}_{d, 2i-1}\right)$, and $\mathbf{h}_{2i}= \left(\mathbf{G} \Theta_{2i} \mathbf{h}_{ris, 2i)}+\mathbf{h}_{d, 2i}\right)$. Furthermore, the total achievable rates of $UE_{2i-1}$ and $UE_{2i}$ are the combination of both common and private parts and mathematically can be formulated as:

\begin{equation}
R_{2 i-1}=R_c+ R_{p ,2i-1} ,
\end{equation}

\begin{equation}
R_{2 i}=R_c+ R_{p ,2i} ,
\end{equation}

Consequently, the sum rate can be calculated as SR = $R_{2 i-1}+R_{2 i}$.

\subsection{Problem Formulation}
Based on the expressions in (5) and (6) along with the sum rate, the problem can be formulated to determine the optimal number of RIS elements and the optimal transmit power allocation for the common and private parts. The objective is to maximise the sum rate while ensuring that the required target common rate is achieved:

\begin{subequations} \label{eq7}
	\begin{align}
 \quad & \max_{\Theta, \alpha} \quad (R_{2i-1} + R_{2i}) \label{7a}\\
& \text{s.t.} \quad R_c \geq R_T, \label{7b}\\
& \quad \quad (\alpha_c + \alpha_{p, 2i-i} + \alpha_{p, 2i} )P_T = P_T, \quad \text{where}, \tag{7c}\\
& \quad \quad \quad \alpha_c\geq0, \alpha_{p,k}\geq0, \nonumber\\
& N_{2i-1} + N_{2i} = N, \tag{7d}\\
& \quad \quad N_k \geq N_{min}, \nonumber \\
& \quad \quad |\Theta_k| = 1, \nonumber\\
& 0 \leq \theta_j \leq 2\pi, \quad i \in \{1, \ldots, N_k\}. \tag{7e}
\end{align}
\end{subequations}

The optimization function in (7a) is to maximise the sum rate of the system. (7b) sets the QoS constraint, which states that the common message rate ($R_c$) must be higher or equal to the specified target rate ($R_T$). (7c) ensures that the total transmit power is divided among the common and private messages appropriately. Then, a constraint (7d) defines the number of RIS elements allocated to all users so that the total number of RIS elements. This ensures that a part of RIS elements is allocated to a user and the total number of RIS elements remains fixed. Furthermore, the reflection amplitude is, without loss of generality, assumed to be unity (i.e. $\left|\boldsymbol{\Theta}_k\right|=1$) and (7e) specifies the phase shifts range for each RIS element.

\section{Optimisation Solution: ORIS-RSMA}
For the solution of the optimising problem defined above, the mentioned constraints (7b)-(7e) must be satisfied to find the optimal values for $\alpha_{c}^*, \alpha_{p}^*, N_{good}^*$ and $N_{worst}^*$ for the maximisation of sum rate in (7a). First of all, to find the optimal power allocation $\alpha_{c}^*, \alpha_{p}^*$, the bisection search-based method \cite{sun2015ergodic} is used while assigning an equal number of RIS element $N_k=N/K$ to both users. Once optimal values for power allocation factors $\alpha_{c}^*, \alpha_{p}^*$ are calculated, the user with the worst common rate is identified and termed as $k_{worst}$ as this is the user whose target common rate is to be achieved. After the worst user $k_{worst}$ and good user $k_{good}$ are identified, the optimal number of RIS elements allocated to a user is calculated using the ORIS procedure of Algorithm 1. The details of these two methods are mentioned in Algorithm 1 (ORIS-RSMA).

The ORIS-RSMA algorithm uses a bisection search algorithm to find optimal power and RIS element allocation, resulting in a logarithmic computational complexity. Thus, the overall complexity of the algorithm is $\mathcal{O}\left(\log_2\left(\frac{N}{\epsilon N_{\text{th}}}\right)\right)$. $\epsilon$ is the stopping tolerance for convergence. $N$ is the total number of RIS elements, and $N_th$ is the minimum number of RIS elements that can be allocated to a user. This logarithmic complexity confirms that the algorithm is highly efficient and suitable for real-time and large-scale RIS-RSMA deployments.

\begin{algorithm} 
	\caption{Optimised RIS-RSMA (ORIS-RSMA)}
	
	\hspace*{\algorithmicindent} \textbf{Input:}~$k$, $R_T$, $P_T$, $N$, $N_k$, $N_{th}$, $\textbf{G}$, $\boldsymbol{\Theta}_k$, $\textbf{h}_{ris, k}$, $\textbf{h}_{d, k}$\\
	\hspace*{\algorithmicindent} \textbf{Output:}~$\alpha_{c}*,$ $\alpha_{p}*$, $N_{good}^*$, $N_{worst}^*$ \\
	\hspace*{\algorithmicindent} \textbf{\emph{Notation}}: Superscript $c$ and $p$ represent the common and private parts where $k \in \{2i-1,2i\}$. \\
	\hspace*{\algorithmicindent} \textbf{Initialization:}~$\alpha_{p,min}$=0, $\alpha_{p,max}$=1,\\ \hspace*{\algorithmicindent} \qquad \qquad \qquad  $N_{good}$=0, $N_{worst}$=$N$.
	\begin{algorithmic}[1]
		\Procedure{OPA}{$R_T$, $P_T$, $\alpha_{p,min}$, $\alpha_{p,max}$}
		\While {$\alpha_{p,max}-\alpha_{p,min} \geq \epsilon$}
		\State Set $\alpha_{p}=\left(\alpha_{p, \min }+\alpha_{p, \max }\right) / 2$
		\State Calculate $R_c$ using $N_k=N/K$ in (2)
		\If {$R_{c} \leq R_{T}$}
		\State $\alpha_{p, \max }=\alpha_{p}$
		\Else
		\State $\alpha_{p, \min }=\alpha_{p}$
		\EndIf
		\EndWhile
		\State \textbf{Output:}~$\alpha_{p}^{*}=\alpha_{p}, \alpha_{c}^{*}=\alpha_{p,\max}-\alpha_{p}^{*} $
		\EndProcedure\\
        \textbf{Identify the better-performing and worst-performing user in terms of user rate:} $k_{good}, k_{worst}$
        
		\Procedure{ORIS}{$R_T$, $\alpha_p^*,$ $\alpha_c^*$, $N_{th}$, $N_{good,min}$, $N_{good,max}$}
        
		\While {$N_{good,max}-N_{good,min} \geq N_{th}$}
		\State Set $N_{good}=\lceil  \left(N_{good, \min }+N_{good, \max }\right) / 2 \rceil$
		\State Calculate $N_{worst}$=$N-N_{good}$
		\State Calculate $\textbf{G}$, $\boldsymbol{\Theta}_k$, $\textbf{h}_{ris, k}$ using $N_{good}, N_{worst}$
		\State Calculate $R_c$ using $\textbf{G}$, $\boldsymbol{\Theta}_k$, $\textbf{h}_{ris,k}$, $\alpha_c^*,$ $\alpha_p^*$ in (2)
		\If {$R_{c} \leq R_{T}$}
		\State $N_{good, \max }=N_{good}$
		\Else
		\State $N_{good, \min }=N_{good}$
		\EndIf
		\EndWhile
		\State \textbf{Output:}~$N_{good}^{*}=N_{good}, N_{worst}^{*}=N_{good,\max}-N_{good}^{*} $
		\EndProcedure
	\end{algorithmic}
\end{algorithm}

\section{Result Analysis}
This section analyses the proposed algorithm's performance compared with the conventional methods in terms of sum rate. The results are averaged after Monte Carlo simulations. The simulation parameters are: total RIS elements $N=256$, number of users $K=2$, total transmit power (Normalized) $P_T=1$, the target rate $R_T= 2bps/Hz$ and transmit SNR 0-30dB. The normalized distances between BS-UE$_{2i}$, BS-UE$_{2i-1}$, BS-RIS, RIS-UE$_{2i}$ and RIS-UE$_{2i-1}$ are 0.5, 1, 0.4, 0.3, 0.6,  respectively. For conventional RSMA and RIS-RSMA, the fixed power allocation (FPA) is employed where for common part $\alpha_c=0.5$ and private part $\alpha_p= (1-\alpha_c)/2$ and the number of RIS elements allocated to each user in RIS-RSMA $N_k=128$.
\begin{figure}[t!]
	\centerline{\includegraphics[scale=0.6]{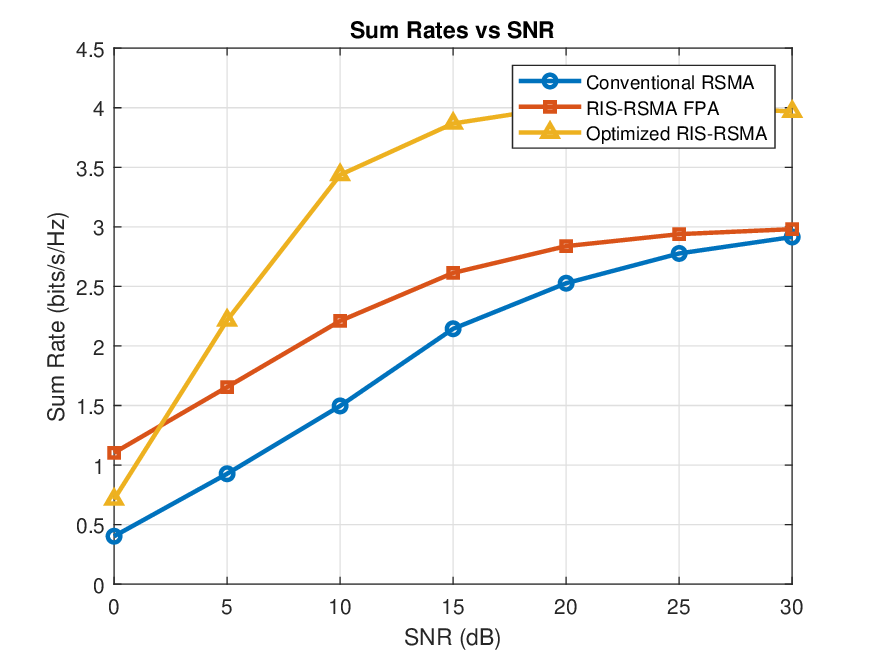}}
	\caption{Sum rates comparison of ORIS-RSMA, RIS-RSMA and conventional RSMA}
	\label{fig2}
\end{figure}
Fig. 2 shows the sum rate comparison of ORIS-RSMA, RIS-RSMA and conventional RSMA. The conventional RSMA is used as a baseline, and it shows a low sum rate for all SNR values as expected. With the introduction of RIS, the transmission channel improves and the sum rate also increases. However, it can be seen that the sum rate for ORIS-RSMA is higher than the other schemes, particularly at low and medium SNR values as expected due to the power and RIS elements optimisation. At low SNRs, the improvement is significant due to the fact that at low SNRs, power constraints are more limiting, and ORIS-RSMA is balancing the resources effectively. In addition at high SNR values, the improvement is minimal as the system's dependence on resource optimisation reduces.

The results in Fig. 3 demonstrate the optimal number of RIS elements to the users along with power allocation to the common part in ORIS-RSMA. At low SNR, more RIS elements are allocated to better-performing user $N_{good}$ to maximise the sum rate by improving the channel. This effectively utilises the additional RIS elements as well as more power to the common part so that all users can decode the common part successfully. At higher SNR values, more RIS elements are allocated to the worst user to ensure fairness. Additionally, the power allocation for the common stream is reduced to allocate more power to the private parts as at higher SNR, they rely more on private streams without compromising the common rate constraint.

\begin{figure}[t!]
	\centerline{\includegraphics[scale=0.6]{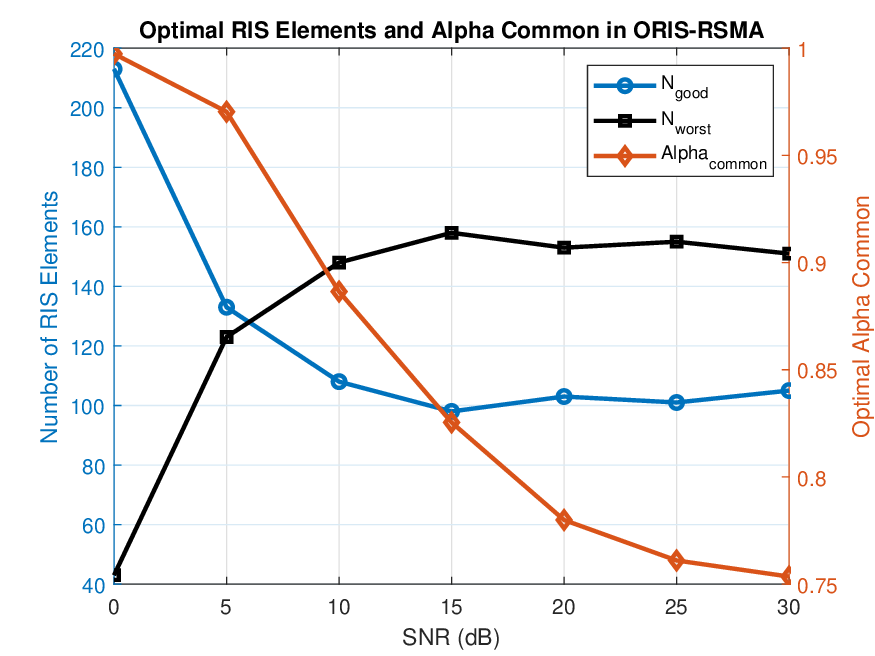}}
	\caption{Optimal RIS elements and power allocation in ORIS-RSMA}
	\label{fig3}
\end{figure}

\section{Conclusion}
In this article, the power allocation factors for RSMA and the number of RIS elements allocated to a user in a RIS-RSMA group are optimized while satisfying the minimum target rate for the common stream to maximise the overall sum rate. For this purpose, an optimization (ORIS-RSMA) algorithm based on bisection search optimization is proposed to obtain the optimal power allocation and number of RIS elements. Lastly, the proposed ORIS-RSMA results are compared with RIS-RSMA and conventional RSMA, to prove its superiority in terms of sum rate via simulations. In future, the system can be analysed for imperfect CSI and MIMO cases.

\bibliographystyle{IEEEtran}
\bibliography{bibliography}
\end{document}